\documentclass[11pt,a4paper]{article}
\usepackage[hyperref]{acl2019}
\usepackage{times}
\usepackage{latexsym}
\usepackage{graphicx}
\usepackage{amsmath}

\usepackage{url}

\aclfinalcopy % Uncomment this line for the final submission
\title{Collaborative Multi-Agent Dialogue Model Training Via Reinforcement Learning}

\author{
  Alexandros Papangelis, Yi-Chia Wang, Piero Molino, Gokhan Tur \\
  Uber AI \\
  San Francisco, California \\
  {\tt \{apapangelis, yichia.wang, piero, gokhan\}@uber.com} \\
}

\date{}

\begin{document}
\maketitle
\begin{abstract}
We present the first complete attempt at concurrently training conversational agents that communicate only via self-generated language. Using DSTC2 as seed data, we trained natural language understanding (NLU) and generation (NLG) networks for each agent and let the agents interact online. We model the interaction as a stochastic collaborative game where each agent (player) has a role (``assistant'', ``tourist'', ``eater'', etc.) and their own objectives, and can only interact via natural language they generate. Each agent, therefore, needs to learn to operate optimally in an environment with multiple sources of uncertainty (its own NLU and NLG, the other agent's NLU, Policy, and NLG). In our evaluation, we show that the stochastic-game agents outperform deep learning based supervised baselines.
\end{abstract}

\section{Introduction}
Machine learning for conversational agents has seen great advances \citep[e.g.][]{SLUBook, INR-074, singh2000reinforcement, young2013pomdp, oh2000stochastic, zen2009statistical, reiter2000building, rieser2010natural}, especially when adopting deep learning models \cite{dlnlpbook, mesnil2015using, wen2015semantically, wen2016network, su2017sample, papangelis2018spoken, liu2018end, li2017end, williams2017hybrid, liu2018adversarial}. 
%van2016wavenet,
Most of these works, however, suffer from the lack of data availability as it is very challenging to design sample-efficient learning algorithms for problems as complex as training agents capable of meaningful conversations. Among other simplifications, this results in treating the interaction as a single-agent learning problem, i.e. assuming that from the conversational agent's perspective the world may be complex but is stationary.  In this work, we model conversational interaction as a stochastic game \citep[e.g.][]{bowling2000analysis} and train two conversational agents, each with a different role, which learn by interacting with each other via natural language. We first train Language Understanding (NLU) and Generation (NLG) neural networks for each agent and then use multi-agent reinforcement learning, namely the Win or Lose Fast Policy Hill Climbing (WoLF-PHC) algorithm \cite{bowling2001rational}, to learn optimal dialogue policies in the presence of high levels of uncertainty that originate from each agent's statistical NLU and NLG, and the other agent's erratic behaviour (as the other agent is learning at the same time). While not completely alleviating the need for seed data needed to train the NLU and NLG components, the multi-agent setup has the effect of augmenting them, allowing us to generate dialogues and behaviours not present in the original data. 

\begin{figure*}[t]
\includegraphics[width=\linewidth]{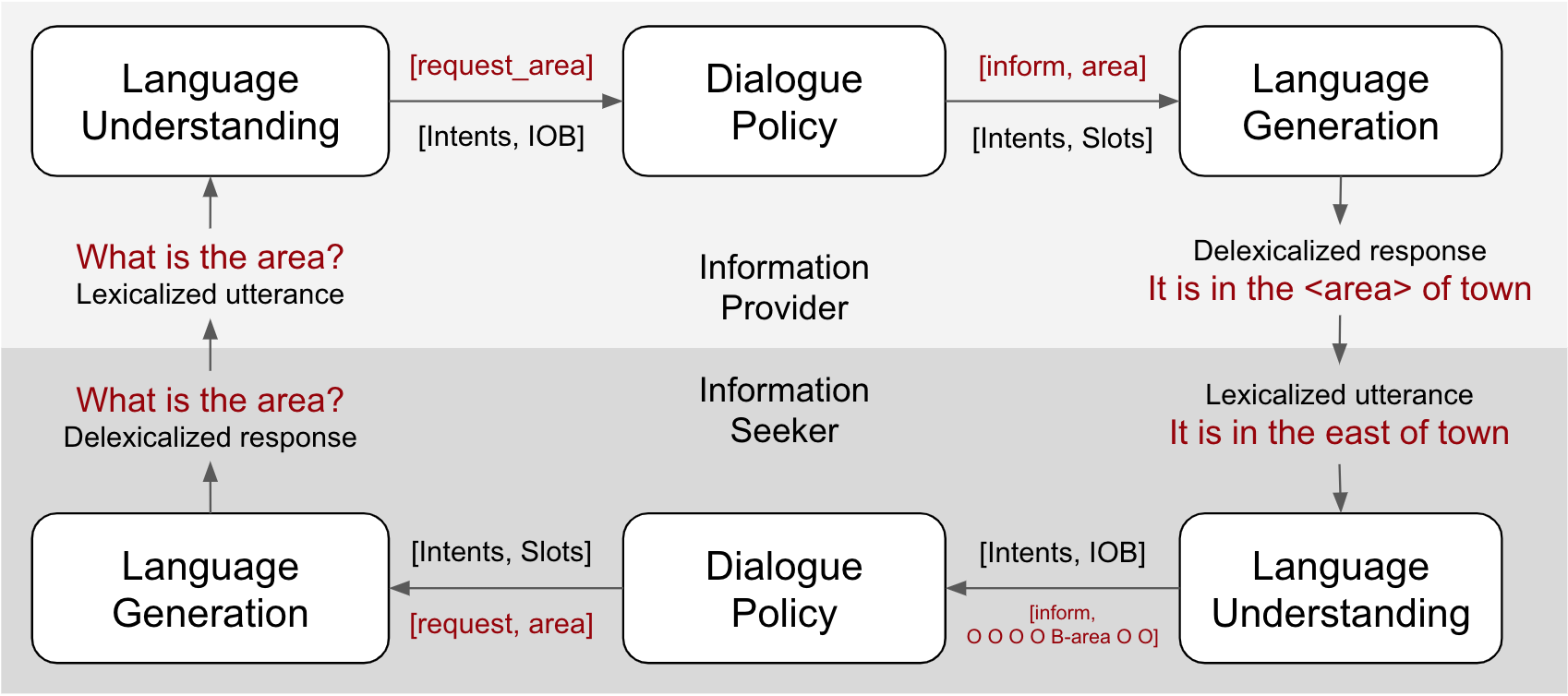}
\centering
\caption{Information flow between two agents on a successful example (shown in red, starting from the Information Seeker's policy).
%; in general, errors may occur at any point.
Where needed, slot values are populated from the tracked dialogue state.}
\label{fig:sysarch}
\end{figure*}

Employing a user simulator is an established method for dialogue policy learning~\cite[among others]{schatzmann2007agenda} and end-to-end dialogue training~\cite{asri16, liu2018end}. Training two conversational agents concurrently has been proposed by \citet{georgila2014single}; training them via natural language communication was partially realized by \citet{liu2017iterative}, as they train agents that receive text input but generate dialogue acts. However, to the best of our knowledge, this is the first study that allows fully-trained agents to communicate only in natural language, and does not allow any all-seeing critic / discriminator. Inspired by \citet{dileknaacl18}, each agent learns in a decentralized setting, only observing the other agent's language output and a reward signal. This allows new, untrained agents to directly interact with trained agents and learn without the need for adjusting parameters that can affect the already trained agents. 

The architecture of each agent is mirrored as shown in Figure~\ref{fig:sysarch}, so the effort of adding agents with new roles is minimal. As seed data, we use data from DSTC2 \cite{henderson2014second}, which concerns dialogues between humans asking for restaurant information and a machine providing such information. Our contributions are: 1) we propose a method for training  fully text-to-text conversational agents from mutually generated data; and 2) we show how agents trained by multi-agent reinforcement learning and minimal seed human-machine data can produce high quality dialogues as compared to single-agent policy models in an empirical evaluation.

%, and 3) we compare against established methods for NLU, NLG, and dialogue policy learning.

\begin{table*}[t]
    \centering
    \begin{tabular}{|c|l|}
        \hline
        Goal & Constr(pricerange=cheap), Constr(area=north), Req(addr), Req(phone) \\ \hline
        {\bf Agent Role} & {\bf Input / Output}\\ \hline
        & \emph{Example of DM error (Seeker's policy is also learning):} \\ \hline
        Prov. NLG & what part of town do you have in mind?  \\ 
        Seeker NLU & request({\bf area}) \\
        Seeker DM & act\_inform {\bf food}  \\ \hline
        & \emph{Example of NLG error:} \\ \hline
        Seeker DM & act\_request phone \\
        Seeker NLG & what is the phone \\ 
        Prov. NLU & request(phone) \\
        Prov. DM & act\_inform {\bf phone} \\
        Prov. NLG & the {\bf post code} is c.b 4, 1 u.y . \\
        Seeker NLU & inform(postcode = c.b 4, 1 u.y)\\ \hline
        & \emph{Example of NLU error:} \\ \hline
        Provider NLG & the phone number is {\bf 01223 356555} \\ 
        Seeker NLU & inform(phone={\bf 01223}) \\ \hline
        
    \end{tabular}
    \caption{A failed dialogue between two conversational agents during training. Uncertainty originating from NLU and NLG components on top of the erratic behaviour of each agent's policy (as they learn concurrently) can have a big impact on the quality of the learned dialogue policies.}
    \label{tab:example}
    \vspace{-2mm}
\end{table*}

\subsection{Related Work}
\label{ssec:rw}
Collecting and annotating a big corpus requires significant effort and has the additional challenge that agents trained in a supervised manner with a given corpus cannot easily generalize to unseen / out of domain input. Building a good user simulator to train against can be challenging as well, even equivalent to building a dialogue system in some cases. Directly learning from humans leads to policies of higher quality, but requires thousands of dialogues even for small domains \cite{gavsic2013line}. \citet{shah2018bootstrapping} combine such resources to train dialogue policies. Recently, model-based RL approaches to dialogue policy learning are being revisited \cite{wu2018switch}; however, such methods still assume a stationary environment. 

\citet{georgila2014single} concurrently learn two negotiator agents' dialogue policies in a setting where they negotiate allocation of resources. However, their agents do not interact via language, but rather via dialogue acts. They use PHC and WoLF-PHC \cite{bowling2001rational} to train their agents, who use two types of dialogue acts: \textit{accept} and \textit{offer}, each of which takes two numerical arguments. \citet{Lewis:2017} train agents on a similar task, but their agents are modelled as end-to-end networks that learn directly from text. However, the authors train their negotiator agent on supervised data and against a fixed supervised agent. Earlier works include \citet{english2005learning}, the first to train policies for two conversational agents, but with single-agent RL, and \citet{chandramohan2014co} who applied co-adaptation on single-agent RL, using Inverse RL to infer reward functions from data. 

\citet{liu2017iterative} train two agents on DSTC2 data, taking text as input and producing dialogue acts that are then fed to template-based language generators. They pre-train their models using the data in a supervised manner and apply reinforcement learning on top. In our setup, information providers and seekers are modeled as active players in a non-stationary environment who interact with each other via language they generate, using statistical language generators. Each agent has their own reward as the objectives are not identical, and their dialogue manager uses a method designed for non-stationary environments. While our setup still needs seed data to ensure linguistic consistency and variability, it augments this data and can train high quality conversational agents.

Other than the works mentioned above, many approaches have been proposed to train modular or end-to-end dialogue systems. To the best of our knowledge, however, none of them concurrently trains two conversational agents.

\section{System Overview}
\label{sec:system}

Figure~\ref{fig:sysarch} shows the general architecture and information flow of our system, composed of two agents who communicate via written language. Our system operates in the well-known DSTC2 domain \cite{henderson2014second} which concerns information about restaurants in Cambridge; however, our multi-agent system supports any slot-filling / information-seeking domain. The Language Understanding and Generation components are trained offline as described in the following sections, while the dialogue policies of the agents are trained online during their interaction. Given that our language generation component is model-based rather than retrieval-based or template-based, we believe that the quality of the generated language and dialogues is encouraging (see appendix for some example dialogues).

\subsection{Language Understanding}
\label{sec:lu}

The task of Natural Language Understanding (NLU) consists of mapping a free-form sentence to a meaning representation, usually in the form of a semantic frame. The frame consists of an intent and a set of slots with associated values. For instance, the semantic frame of the sentence \emph{``Book me an Italian restaurant in the south part of the city''} can be mapped to the frame \emph{``book\_restaurant (food: Italian, area: south)''} where \emph{book\_restaurant} is the intent and \emph{food} and \emph{area} are the slots.
%Dialogue systems that deal with other types of tasks usually add more levels of abstraction that model intents more granularly.

In recent years, deep learning approaches have been adopted for NLU, performing intent classification and slot tagging both independently \cite{Tr2012TowardsDU, Lee2016SequentialSC, Xu2013ConvolutionalNN, mesnil2015using, Kurata2016LeveragingSI,Huang2015BidirectionalLM}
and jointly \cite{Zhang2016AJM, RojasBarahona2016ExploitingSA}. In \citet{HakkaniTr2016MultiDomainJS}, decoders tag each word in the input sentence with a different slot name and concatenate the intent as a tag to the end-of-sentence token, while in \citet{Liu2016AttentionBasedRN} the encoder is shared, but the two tasks have separate decoders. In most cases, intent detection is treated as a classification problem and the slot name tags for all words are uniquely assigned to the intent detected in the sentence.

In our case, as we decided to use the same NLU model architecture for both agent roles, we could not rely on multi-class classification. In particular, system outputs in DSTC2 often contain multiple acts, so an ``information seeker" NLU model has to learn to identify which intents are present in the system utterance as well as to assign slot values to each identified intent. An example of this need is evident in the sentence \emph{``There are no Italian restaurants in the south part of the city, but one is available in the west side''} which can be mapped to \emph{``\{deny(food: Italian, area: south), inform(area: west)\}"}. In order to tackle those scenarios, we designed our decoder to predict multiple intents (casting the task as a multi-label classification problem) where each intent is a class and, for the \emph{``request''} intent, the pair of \emph{``request''} and all requestable slots are additional classes. This is necessary as the slot values of the request intent are names of slots (e.g. \emph{request(food)}), and they may not be mentioned explicitly in the sentences. Moreover, to account for the multiple intents in the set tagger decoder, we augmented the number of possible tags for each word in the sentence concatenating the name of the intent they are associated with. In the previous example, for instance, the word ``south'' is assigned a \emph{``deny\_area''} tag, while the word ``west'' is assigned an \emph{``inform\_area''} tag, so the name of the intent in the tag identifies which of the multiple intents each slot is assigned to. This increases the number of tags, but allows an unequivocal assignment of the slot values to the intents they belong to.

The whole model, which is composed of a convolutional encoder and the two decoders (one intent multi-label classifier and a slot tagger), is trained end-to-end in a multi-task fashion, with both multi-label intent classification and slot tagging tasks being optimized at the same time. The output set of semantic frames from the NLU is then aggregated over time and passed on to the dialogue policy.

\paragraph{Evaluating NLU Quality}
% As both provider NLU and seeker NLU return sets of intents and frames, we decided to use the Jaccard index as a way to measure the overlap between the ground truth sets and the sets predicted by each model. Intent average Jaccard is the average of the the Jaccard index of the sets of intents for each datapoint, frame average Jaccard is the average of the the Jaccard index of the sets of frames for each datapoint (two frames match if the intents and the values for every slot match) and slot average F1 is the standard F1 measure for the slot values. 

Table \ref{tab:lu_results} summarizes the performance (F1 scores) of the trained models, with respect to intent, frame, and slot IOB tags, calculated on the DSTC2 test set.
The F1 measure is used instead of accuracy due to the multiple intents, acts and slots in our problem formulation.

\begin{table}[h]
    \centering
    \begin{tabular}{|r|c|c|c|}
        \hline
        % Model & Int. Jacc & Frame Jacc & \\ \hline
        % Prv. & 0.917 & 0.914 & \\
        % Skr. & 0.984 & 0.981 & \\ \hline
        Role & Intent F1 & Slots F1 & Frame F1 \\ \hline
        Provider & 0.929 & 0.899 & 0.927 \\
        Seeker & 0.986 & 0.995 & 0.983 \\ \hline
    \end{tabular}
    \caption{F1 scores for each agent's NLU model.}
    \label{tab:lu_results}
    \vspace{-5mm}
\end{table}

\subsection{Dialogue Policy Learning}
\label{sec:dpl}
As already discussed, in this work we train two agents: one seeking restaurant information (``seeker") and one providing information (``provider"). Each agent's dialogue policy receives the tracked dialogue state and outputs a dialogue act. While both agents have the same set of dialogue acts to choose from, they have different arguments to use for these acts \cite{henderson2014second}. Each agent also has a different dialogue state, representing its perception of the world. The seeker's state models its preferences (goal) and what information the provider has given, while the provider's state models constraints expressed or information requested by the seeker, as well as attributes of the current item in focus (retrieved from a database) and metrics related to current database results, such as number of items retrieved, slot value entropies, etc. The reward signal is slightly different for each agent, even though the task is collaborative. It assigns a positive value on successful task completion (restaurant provided matches the seeker's goal, and all seeker's requests are answered), a negative value otherwise, and a small negative value for each dialogue turn to favor shorter interactions. However, a seeker is penalised for each request in the goal that is not expressed, and a provider is penalised for each request that is unanswered. To train good dialogue policies in this noisy multi-agent environment, we opted for WoLF-PHC as a proof of concept and leave investigation of general-sum and other methods that scale better on richer domains for future work. The dialogue policies that we train operate on the full DSTC2 act and a subset of the slot space. Specifically, not all dialogue acts have slot arguments and we do not allow multiple arguments per act or multiple acts per turn, so the size of our action space is 23. In the input, all policies receive the output of the NLU aggregated over the past dialogue turns (i.e. keeping track of slots mentioned in the past) with - as mentioned above - the state of the seeker including its own goal, and the state of the provider including current database result metrics which are fetched through SQL queries formed using the slot-value pairs in the provider's state.

\subsubsection{WoLF-PHC}
\label{sssec:marl}
A \emph{stochastic game} can be thought of as a \emph{Markov Decision Process} extended to multiple agents. It is defined as a tuple $(n, S, A_{1..n}, T, R_{1..n})$, where $n$ is the number of agents, $S$ is the set of states, $A_i$ is the set of actions available to agent $i$, $T:S \times A \times S \rightarrow [0,1]$ is the transition function, and $R_i:S\times A \rightarrow \Re$ is the reward function of agent $i$.

WoLF-PHC \cite{bowling2001rational} is a PHC algorithm (simple extension to Q-Learning for mixed policies) with variable learning rate and the principle according to which the agent should learn quickly (i.e. with a higher learning rate) when losing and slowly when winning. Briefly, $Q$ is updated as in Q-Learning and an estimate of the average policy is maintained:\\ $\tilde{\pi}(s, a^\prime) \leftarrow \tilde{\pi}(s, a^\prime) + \frac{1}{C(s)}(\pi(s, a^\prime) - \tilde{\pi}(s, a^\prime))$, where $C(s)$ is the number of times state $s$ has been visited. The policy then is updated as follows: 

\begin{equation*}
    \pi(s, a) \leftarrow \pi(s, a) + 
    \begin{cases}
      \delta & a = amax_{a^\prime}Q(s, a^\prime) \\
      \frac{-\delta}{|A_i|-1} & otherwise
    \end{cases}
\end{equation*}

\begin{equation*}
    \delta = 
    \begin{cases}
        \delta_w & \sum_a \pi(s,a)Q(s,a) > \sum_a \tilde{\pi}(s,a)Q(s,a) \\
        \delta_l & otherwise
    \end{cases}
\end{equation*}\\
where $\delta_w$ and $\delta_l$ are learning rates.

% MinimaxQ \cite{littman1994markov} is a logical extension to Q-learning \cite{}, where the main concept is that the Q function is extended to model the other agents' actions: $Q:S\times A_i \times A_j \rightarrow \Re$ and each agent chooses an action on the assumption that the other agent will act optimally. The update of $Q$ is defined as follows: $Q(s,a_i,a_j) \leftarrow (1-\alpha)Q(s,a_i,a_j) + \alpha (R(s,a_i) + \gamma V(s^\prime))$, where $\alpha$ is the learning rate. A linear program then needs to be solved when updating the policy: $\pi(s,\cdot) \leftarrow argmax_{\pi^\prime(s,\cdot)} min_{a_j} \sum_{a_i^\prime} \pi^\prime(s, a_i^\prime)Q(s, a_i^\prime, a_j)$. The value function then is updated as: $V(s) \leftarrow min_{a_j} \sum_{a_i^\prime} \pi(s, a_i^\prime)Q(s,a_i^\prime,a_j^\prime)$.

% This makes MinimaxQ not very efficient or scalable but it fits our proof-of-concept purposes. As the authors note, in the case of games of alternating turns (similar to a conversation) linear programming is not necessary to compute the value function as in that case there is an optimal deterministic policy and $V(s) = max_{a_i} min_{a_j} Q(s,a_i,a_j)$.

\subsection{Language Generation}
\label{sec:lg}

% Definition of nlg, The input and output of nlg
Natural language generation (NLG) is a critical module in dialogue systems.  It operates in the later phase of the dialogue system, consumes the meaning representation of the intended output provided by the dialogue manager, and converts it to a natural language utterance. 

% Brief literature overview of common nlg approaches
Previous research has approached the NLG problem in various ways \citep[e.g.,][]{langkilde1998generation, walker2002training, oh2000stochastic}.  One common approach is rule-based / template-based generation, which produces utterances from handcrafted rules or templates where slot variables are filled with values from the meaning representation provided by the dialogue manager. This approach has been widely adopted in both industrial and research systems. Although it guarantees high-quality output, it is time-consuming to write templates especially for all possible meaning representations and the generated sentences quickly become repetitive for the users. Moreover, scalability and maintenance of these templates become concerns as we expand the system to deal with more domains or scenarios.  

% Later, some researchers divided the NLG process into three components (\textit{sentence planner, surface realizer, and prosody assigner}) and developed trainable solutions for them \citep[e.g.,][]{langkilde1998generation, walker2002training}.  Nevertheless, these systems still require a handcrafted decision space beforehand for model training and optimization.  
% As large corpora and data become more accessible recently, \citet{oh2000stochastic} proposed a corpus-based generation method, which opened up a new direction for the NLG problem.  The idea of corpus-based methods involves building a stochastic generator which learns to produce utterances directly from data and a reranker to sort candidate utterances generated by the generator.  However, these methods usually suffer from accuracy issues and do not generalize well.

More recently, deep neural networks have been widely adopted in natural language generation because of their effectiveness. Among all types of deep learning architectures, the sequence-to-sequence  approach (\textit{seq2seq}) has been most widely and successfully adopted for language generation in several tasks as machine translation \citep[e.g.][]{Sutskever:2014}, question answering \citep[e.g.][]{Yin:2016:NGQ:3060832.3061037}, text summarization \citep[e.g.][]{chopra2016abstractive}, and conversational models \citep[e.g.][]{P15-1152, serban2016building}. 
%DBLP:journals/corr/VinyalsL15, 

% our approach    
% Recent state of the art NLG approaches also rely on \textit{seq2seq} methods. 
% In particular, 
Our NLG model is inspired by recent state of the art \textit{seq2seq} models such as \citet{Sutskever:2014} and \citet{wen2015semantically}, that transform one sequence of words to another.  Our \textit{seq2seq} model was constructed to take a meaning representation string as input and generate the corresponding natural language template as output.  Both input and output were delexicalized with slot values replaced by tags, and values are filled in after the template is generated.  An example of input and output of the system NLG is shown below:

\noindent {\small \texttt{\textbf{Input:} act\_inform <food> act\_inform <pricerange> act\_offer <name>}}

\noindent {\small \texttt{\textbf{Output:} <name> is a great restaurant serving <food> food and it is in the <pricerange> price range}}

Specifically, we implemented our Encoder-Decoder model with Long Short-Term Memory (LSTM) recurrent networks.  We employed an attention mechanism~\cite{bahdanau2014neural} to emphasize relevant parts of the input sequence at each step when generating the output sequence. We further improved the model by encoding the conversation history as a context vector and concatenating it with the encoded input for output generation.  We observed that context not only increases the model performance, but also helps to produce output with more \textit{variation}, which has been considered one of the important factors of a good NLG model \cite{stent2005evaluating}. Both agents' NLG models were built in the same way using the provider- or seeker-side data.

\paragraph{Evaluating NLG Quality}

BLEU score \cite{papineni2002bleu} has been one of the most commonly used metrics for NLG evaluation.  Since it is agreed that the existing automatic evaluation metrics for NLG have limitations \cite{belz2006comparing}, we introduced a modified version of BLEU which attempts to compensate the gap of the current BLEU metric.  BLEU, ranging from 0 to 1, is a precision metric that quantifies n-gram overlaps between a generated text and the ground truth text.  However, we observed that in the DSTC2 data a meaning representation can map to different templates as the example shown below:

\noindent {\small \texttt{\textbf{MR:} act\_inform <pricerange> act\_offer <name>}}

\noindent {\small \texttt{\textbf{T1:} the price range at <name> is <pricerange>}}

\noindent {\small \texttt{\textbf{T2:} <name> is in the <pricerange> price range}}

Thus, to compute BLEU of a model-generated template, instead of only comparing it against its corresponding ground truth template, we calculated its BLEU scores with all the possible templates that have the same input meaning representation in the DSTC2 data, and the maximum BLEU score among them is the final BLEU of this generated template.  By doing so, the average BLEU scores of the information provider and seeker NLG models on the test set are 0.8625 and 0.5293, respectively.  Note that it is not surprising that the seeker model does not perform as well as the provider model because the seeker-side data has many more unique meaning representations and natural language templates, which make the task of building a good seeker model harder.

\section{Evaluation}
\label{sec:eval}
The Plato Research Dialogue System\footnote{The source code for the full dialogue system can be found here \url{https://github.com/uber-research/plato-research-dialogue-system}} was used to implement, train, and evaluate the agents. To assess the quality of the dialogues our agents are capable of, we compare dialogue success rates, average cumulative rewards, and average dialogue turns along two dimensions: a) access to ground truth labels during training or not; b) stationary or non-stationary environment during training. We therefore train four kinds of conversational agents for each role (eight in total) as shown in Table \ref{tab:agents}. Due to the nature of our setup, algorithms designed for stationary environments (e.g. DQN) are not considered.

\begin{table}[h]
    \centering
    \begin{tabular}{|r|c|c|}
        \hline
        & Stat. Env. & Non-Stat. Env. \\ \hline
        Dial. Acts & SuperDAct & WoLF-Dact \\
        Text & Supervised & WoLF-PHC \\ \hline
    \end{tabular}
    \caption{The four conditions under which our conversational agents are trained.}
    \label{tab:agents}
    \vspace{-3mm}
\end{table}

\begin{figure}[t]
\includegraphics[width=1.0\linewidth]{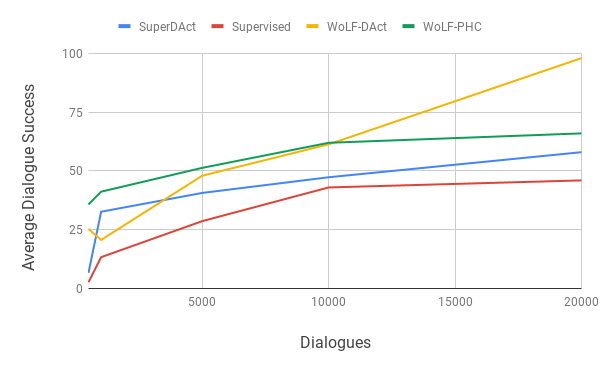}\\
\includegraphics[width=1.0\linewidth]{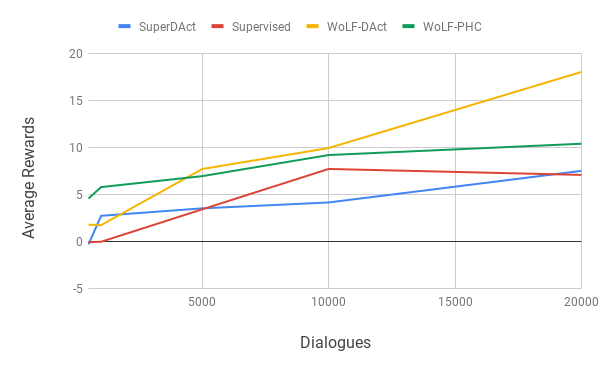}\\
\includegraphics[width=1.0\linewidth]{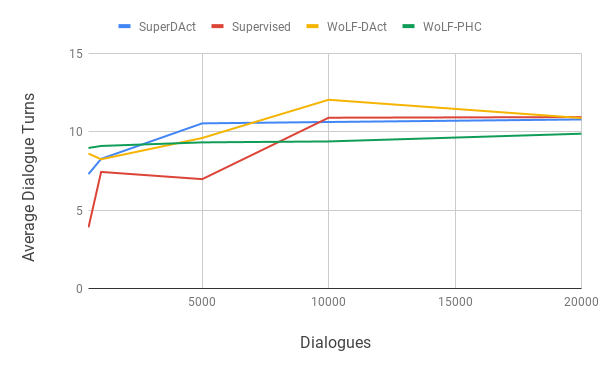}
\centering
\caption{Learning curves of the dialogue policies of our conversational agents, each evaluated on the environment it is trained on (see Table \ref{tab:agents}). Note that the agents are evaluated against each other, not against rational simulators or data.}
\label{fig:learningCurves}
\vspace{-5mm}
\end{figure}

Specifically, the \emph{SuperDAct} agents are modelled as 3-layer Feed Forward Networks (FFN), trained on DSTC2 data using the provided dialogue act annotations. The \emph{Supervised} agents (also 3-layer FFN) are trained on DSTC2 data but each agent's policy uses the output of its respective NLU: the provider (dialogue system in the dataset) generates its utterance using its trained NLG with the dialogue acts found in the data as input; the seeker (human caller in the dataset) then uses the provider's  utterance as input to its NLU whose output is then fed to its policy; and the same approach is used for the provider's side. The \emph{WoLF-DAct} agents are trained concurrently (i.e. in a non-stationary environment) but interacting via dialogue acts, while the \emph{WoLF-PHC} agents are trained concurrently and interacting via generated language, as show in in Figure \ref{fig:sysarch}. All of these agents are then evaluated on the full language to language setup \footnote{The \emph{SuperDAct} and \emph{WoLF-DAct} agents achieve 81\% and 95\% dialogue success rates respectively when evaluated on a dialogue act to dialogue act setup (i.e. without LU/LG) against an agenda-based simulated Seeker. When evaluated against each other (Fig. \ref{fig:learningCurves}) the performance naturally drops.}. Apart from the above, we trained conversational agents using deep policy gradient algorithms. Their performance could not match the WoLF-PHC or the supervised agents, however, even after alternating the policy gradient agents' training to account for non-stationarity. This is not unexpected, of course, since those algorithms are designed to learn in a stationary environment. These results therefore are not reported here.

In our evaluation, a dialogue is considered successful if the information seeker's goal is met by the provider, following the standard definition used for this domain \cite[e.g.]{su2017sample}. Under this definition, a provider must offer an item that matches the seeker's constraints and must answer all requests made by the seeker. However, as seen in Table \ref{tab:example}, even when the dialogue manager's output is correct, it can be realized by NLG or understood by NLU erroneously. While none of the models (NLU, DM, NLG) directly optimises this objective, it is a good proxy of overall system performance and allows for direct comparison with prior work. As a reward signal for reinforcement learning we use the standard reward function found in the literature \cite[e.g.]{gavsic2013line, su2017sample}, tweaked to fit each agent's perception as described in section \ref{sec:dpl}.

\begin{table*}
    \centering
    \begin{tabular}{|c|c|c|c|}
        \hline
        \multicolumn{4}{|c|}{{\bf Average Dialogue Success}} \\
        \hline
        SuperDAct & Supervised & WoLF-DAct & WoLF-PHC \\
        \hline
        44.23\% & 46.30\% & 52.56\% & {\bf 66.30\%} \\ \hline
        \multicolumn{4}{|c|}{{\bf Average Cumulative Rewards}} \\
        \hline
        SuperDAct & Supervised & WoLF-DAct & WoLF-PHC \\
        \hline
        4.42 & 6.68 & 7.84 & {\bf 10.93} \\ \hline
        \multicolumn{4}{|c|}{{\bf Average Dialogue Turns}} \\
        \hline
        SuperDAct & Supervised & WoLF-DAct & WoLF-PHC \\
        \hline
        10.89 & {\bf 8.65} & 9.81 & 9.57 \\ \hline
    \end{tabular}
    \caption{Average dialogue success, reward, and number of turns on the agents evaluated, over 3 training/evaluation cycles with goals sampled from the test set of DSTC2. Regardless of training condition, all agents were evaluated in the language to language setting. All differences between SuperDAct - WoLF-DAct, and Supervised - WoLF-PHC are significant with p $<$ 0.02.}
    \label{tab:results}
\end{table*}

Figure \ref{fig:learningCurves} shows learning curves with respect to the metrics we use for all conversational agents, where each kind of agent was evaluated against its counterpart (e.g. \emph{Supervised} seeker against \emph{Supervised} provider) on the environment they were trained on. Table \ref{tab:results} shows the main results of our evaluation in the language to language setting, where each cell represents the average of 3 train/evaluation cycles of policies trained under the respective conditions for 20,000 dialogues (200 epochs for the supervised agents) and evaluated for 1,000 dialogues. We can see that the WoLF-PHC agents outperform the other conditions in almost every metric, most likely because they model the conversation as a stochastic game and not as a single-agent problem. Comparing Figure \ref{fig:learningCurves} with Table \ref{tab:results} we can see that the agents trained on dialogue acts cannot generalise to the language to language setting, even when paired with NLU and NLG models that show strong performance (see previous section). On a similar setup (joint NLU and DM but without statistical NLG), \citet{liu2017iterative} report 35.3\% dialogue success rate for their supervised baseline and 64.7\% for reinforcement learning on top of pre-trained supervised agents. 

We attribute the low performance of the supervised policies to a lack of data and context in the DSTC2 dataset. We believe that in the presence of errors from our statistical NLU and NLG, there just are not enough dialogues or information within each dialogue for the supervised policies to learn to associate states with optimal actions. In particular, if one of the NLGs or NLUs (for either agent) makes a mistake, this affects the dialogue state tracking and subsequently the database retrieval, resulting in a state that may not actually be in the dataset. In the presence of this uncertainty we found that seeker and provider do not properly learn how to make requests and address them, respectively and this is the most frequent reason for dialogue task failure in this condition. This is partly due to the fact that in DSTC2 the provider's side responds to requests with an offer and an inform, for example a response to a request for phone number would be: offer(name=kymmoy), inform(phone=01223 311911) which may be confusing both models. In light of this, we trained a supervised policy model able to output multiple actions at each dialogue turn. However, this makes the learning problem even harder and we found that in this case such models perform poorly. Overall the two supervised approaches appear to perform similarly on objective dialogue task success but the \emph{Supervised} agents who have seen uncertainty during the training seem to perform better in terms of rewards achieved and number of dialogue turns.

Upon pairing different combinations of the eight agents we trained, we observe that agents who are able to better model the seeker's behaviour perform best in the joint task. In our case, WoLF-trained agents are able to better model the seeker's behaviour, which partially explains the higher success rates. However, we note that the \emph{WoLF-DAct} agents do not generalise very well to the much harder language to language environment. Another general trend that we observe is that the WoLF-trained agents seem to take longer number of turns but lead to higher rewards and success rates likely because they persist for more turns before giving up.

It is also worth noting that while we report an objective measure of dialogue success (i.e. if both agents achieved the goal), from each agent's perspective what is success may be different. For example, if a seeker does not inform about all constraints in the goal but provider respects all constraints that the provider does mention then the dialogue is successful from the provider's perspective but failed from the seeker's perspective. On the other hand, if the seeker provides all constraints and requests but the provider either ignores some constraints, says it cannot help, or does not address some requests then the dialogue is failed from the provider's perspective but successful from the seeker's perspective. To test whether optimizing the dialogue policies directly against these subjective measures of task success would lead to better dialogue policies, we performed similar experiments as the ones whose results  are reported in Table \ref{tab:results}. However, we found that the overall performance was not as good because it would lead to behaviours in which the agents would not help each other to achieve the objective goal (e.g. the provider would not make many requests, or the seeker would not repeat informs upon wrong offers).

\section{Conclusion}
\label{sec:concl}
We presented the first complete attempt at concurrently training conversational agents that communicate only via self-generated language. Using DSTC2 as seed data, we trained NLU and NLG networks for each agent and let the agents interact and learn online optimal dialogue policies depending on their role (seeker or provider). Future directions include investigating joint optimization of the modules and training the agents online using deep multi-agent RL (e.g. \cite{foerster2018counterfactual}) as well as evaluating our agents on harder environments (e.g. TextWorld \cite{cote2018textworld}) and against human players. A natural extension is to train a multi-tasking provider agent that can learn to serve various kinds of seeker agents.

%[DRAFT] End to end is future work for us - Bing's results show that e2e is not necessarily the solution (other works on e2e also train their models against stationary corpora or simulators whose behaviour does not change). E.g. it would be equally hard (or harder) for an e2e model to learn to associate features from a sentence or dialogue history with specific outputs if its opponent is behaving erratically and the only signal is a numerical reward. So, we will try methods such as COMA, M3DPP, etc that are Deep MARL approaches in our modular setup and in an e2e setup.

% [MAYBE INCLUDE] Future: look at potential e2e applications. While we were able to show that wolf performs better, we need to apply these on harder environments (e.g. using ASR, or TextWorld) because otherwise there is a possibility that RL (whether on top of SL or not) just finds a subspace where most of the actions are irrelevant and the problem is easier. If SL does not have access to ASR confidence and other things, it cannot associate actions such as expl-conf, repeat, conf-domain, etc with given states. Therefore RL will just learn to ignore these and operate on a hello-request-offer-inform-bye action space with a simplistic state (ignoring `irrelevant' features). I.e. regarding RL, we do not learn much unless we apply RL in an environment that almost exactly simulates the environment where the data was collected.

\begin{footnotesize}
\bibliography{acl2019}

\begin{thebibliography}{54}
\expandafter\ifx\csname natexlab\endcsname\relax\def\natexlab#1{#1}\fi

\bibitem[{Asri et~al.(2016)Asri, He, and Suleman}]{asri16}
Layla~El Asri, Jing He, and Kaheer Suleman. 2016.
\newblock A sequence-to-sequence model for user simulation in spoken dialogue
  systems.
\newblock \emph{INTERSPEECH}, pages 1151--1155.

\bibitem[{Bahdanau et~al.(2015)Bahdanau, Cho, and Bengio}]{bahdanau2014neural}
Dzmitry Bahdanau, Kyunghyun Cho, and Yoshua Bengio. 2015.
\newblock Neural machine translation by jointly learning to align and
  translate.
\newblock \emph{ICLR}.

\bibitem[{Belz and Reiter(2006)}]{belz2006comparing}
Anja Belz and Ehud Reiter. 2006.
\newblock Comparing automatic and human evaluation of nlg systems.
\newblock In \emph{11th Conference of the European Chapter of the Association
  for Computational Linguistics}.

\bibitem[{Bowling and Veloso(2000)}]{bowling2000analysis}
Michael Bowling and Manuela Veloso. 2000.
\newblock An analysis of stochastic game theory for multiagent reinforcement
  learning.
\newblock Technical report, Carnegie-Mellon Univ Pittsburgh Pa School of
  Computer Science.

\bibitem[{Bowling and Veloso(2001)}]{bowling2001rational}
Michael~H. Bowling and Manuela~M. Veloso. 2001.
\newblock Rational and convergent learning in stochastic games.
\newblock In \emph{IJCAI}, pages 1021--1026. Morgan Kaufmann.

\bibitem[{Chandramohan et~al.(2014)Chandramohan, Geist, Lef{\`{e}}vre, and
  Pietquin}]{chandramohan2014co}
Senthilkumar Chandramohan, Matthieu Geist, Fabrice Lef{\`{e}}vre, and Olivier
  Pietquin. 2014.
\newblock Co-adaptation in spoken dialogue systems.
\newblock In \emph{Natural Interaction with Robots, Knowbots and Smartphones},
  pages 343--353. Springer.

\bibitem[{Chopra et~al.(2016)Chopra, Auli, and Rush}]{chopra2016abstractive}
Sumit Chopra, Michael Auli, and Alexander~M Rush. 2016.
\newblock Abstractive sentence summarization with attentive recurrent neural
  networks.
\newblock In \emph{NAACL}, pages 93--98.

\bibitem[{C{\^o}t{\'e} et~al.(2018)C{\^o}t{\'e}, K{\'a}d{\'a}r, Yuan, Kybartas,
  Barnes, Fine, Moore, Hausknecht, Asri, Adada et~al.}]{cote2018textworld}
Marc-Alexandre C{\^o}t{\'e}, {\'A}kos K{\'a}d{\'a}r, Xingdi Yuan, Ben Kybartas,
  Tavian Barnes, Emery Fine, James Moore, Matthew Hausknecht, Layla~El Asri,
  Mahmoud Adada, et~al. 2018.
\newblock Textworld: A learning environment for text-based games.
\newblock \emph{arXiv preprint arXiv:1806.11532}.

\bibitem[{Deng and Liu(2018)}]{dlnlpbook}
Li~Deng and Yang Liu, editors. 2018.
\newblock \emph{Deep Learning in Natural Language Processing}.
\newblock Springer.

\bibitem[{English and Heeman(2005)}]{english2005learning}
Michael~S English and Peter~A Heeman. 2005.
\newblock Learning mixed initiative dialog strategies by using reinforcement
  learning on both conversants.
\newblock In \emph{HLT-EMNLP}, pages 1011--1018. Association for Computational
  Linguistics.

\bibitem[{Foerster et~al.(2018)Foerster, Farquhar, Afouras, Nardelli, and
  Whiteson}]{foerster2018counterfactual}
Jakob~N. Foerster, Gregory Farquhar, Triantafyllos Afouras, Nantas Nardelli,
  and Shimon Whiteson. 2018.
\newblock Counterfactual multi-agent policy gradients.
\newblock In \emph{AAAI}, pages 2974--2982. {AAAI} Press.

\bibitem[{Gao et~al.(2019)Gao, Galley, and Li}]{INR-074}
Jianfeng Gao, Michel Galley, and Lihong Li. 2019.
\newblock \href {https://doi.org/10.1561/1500000074} {Neural approaches to
  conversational ai}.
\newblock \emph{Foundations and Trends® in Information Retrieval},
  13(2-3):127--298.

\bibitem[{Gasic et~al.(2013)Gasic, Breslin, Henderson, Kim, Szummer, Thomson,
  Tsiakoulis, and Young}]{gavsic2013line}
Milica Gasic, Catherine Breslin, Matthew Henderson, Dongho Kim, Martin Szummer,
  Blaise Thomson, Pirros Tsiakoulis, and Steve~J. Young. 2013.
\newblock On-line policy optimisation of bayesian spoken dialogue systems via
  human interaction.
\newblock In \emph{ICASSP}, pages 8367--8371. {IEEE}.

\bibitem[{Georgila et~al.(2014)Georgila, Nelson, and
  Traum}]{georgila2014single}
Kallirroi Georgila, Claire Nelson, and David Traum. 2014.
\newblock Single-agent vs. multi-agent techniques for concurrent reinforcement
  learning of negotiation dialogue policies.
\newblock In \emph{ACL}, volume~1, pages 500--510.

\bibitem[{Hakkani-T\"ur(2018)}]{dileknaacl18}
Dilek Hakkani-T\"ur. 2018.
\newblock \href {http://naacl2018.org/keynote.html} {Google assistant or my
  assistant? towards personalized situated conversational agents}.
\newblock In \emph{Proceedings of the 2018 Conference of the North American
  Chapter of the Association for Computational Linguistics: Human Language
  Technologies, Plenary Talk}.

\bibitem[{Hakkani-T{\"u}r et~al.(2016)Hakkani-T{\"u}r, Tur, Celikyilmaz, Chen,
  Gao, Deng, and Wang}]{HakkaniTr2016MultiDomainJS}
Dilek~Z. Hakkani-T{\"u}r, Gokhan Tur, Asli Celikyilmaz, Yun-Nung Chen, Jianfeng
  Gao, Li~Deng, and Ye-Yi Wang. 2016.
\newblock Multi-domain joint semantic frame parsing using bi-directional
  rnn-lstm.
\newblock In \emph{INTERSPEECH}.

\bibitem[{Henderson et~al.(2014)Henderson, Thomson, and
  Williams}]{henderson2014second}
Matthew Henderson, Blaise Thomson, and Jason~D Williams. 2014.
\newblock The second dialog state tracking challenge.
\newblock In \emph{SIGDIAL}, pages 263--272.

\bibitem[{Huang et~al.(2015)Huang, Xu, and Yu}]{Huang2015BidirectionalLM}
Zhiheng Huang, Wei Xu, and Kai Yu. 2015.
\newblock Bidirectional lstm-crf models for sequence tagging.
\newblock \emph{CoRR}, abs/1508.01991.

\bibitem[{Kurata et~al.(2016)Kurata, Xiang, Zhou, and
  Yu}]{Kurata2016LeveragingSI}
Gakuto Kurata, Bing Xiang, Bowen Zhou, and Mo~Yu. 2016.
\newblock Leveraging sentence-level information with encoder lstm for semantic
  slot filling.
\newblock In \emph{EMNLP}.

\bibitem[{Langkilde and Knight(1998)}]{langkilde1998generation}
Irene Langkilde and Kevin Knight. 1998.
\newblock Generation that exploits corpus-based statistical knowledge.
\newblock In \emph{COLING-ACL}, pages 704--710.

\bibitem[{Lee and Dernoncourt(2016)}]{Lee2016SequentialSC}
Ji~Young Lee and Franck Dernoncourt. 2016.
\newblock Sequential short-text classification with recurrent and convolutional
  neural networks.
\newblock In \emph{HLT-NAACL}.

\bibitem[{Lewis et~al.(2017)Lewis, Yarats, Dauphin, Parikh, and
  Batra}]{Lewis:2017}
Mike Lewis, Denis Yarats, Yann Dauphin, Devi Parikh, and Dhruv Batra. 2017.
\newblock \href {https://doi.org/10.18653/v1/D17-1259} {Deal or no deal?
  end-to-end learning of negotiation dialogues}.
\newblock In \emph{EMNLP}, pages 2443--2453.

\bibitem[{Li et~al.(2017)Li, Chen, Li, Gao, and {\c{C}}elikyilmaz}]{li2017end}
Xiujun Li, Yun{-}Nung Chen, Lihong Li, Jianfeng Gao, and Asli
  {\c{C}}elikyilmaz. 2017.
\newblock End-to-end task-completion neural dialogue systems.
\newblock In \emph{IJCNLP}, pages 733--743. Asian Federation of Natural
  Language Processing.

\bibitem[{Liu and Lane(2016)}]{Liu2016AttentionBasedRN}
Bing Liu and Ian Lane. 2016.
\newblock Attention-based recurrent neural network models for joint intent
  detection and slot filling.
\newblock In \emph{INTERSPEECH}.

\bibitem[{Liu and Lane(2017)}]{liu2017iterative}
Bing Liu and Ian Lane. 2017.
\newblock Iterative policy learning in end-to-end trainable task-oriented
  neural dialog models.
\newblock In \emph{ASRU}, pages 482--489. {IEEE}.

\bibitem[{Liu and Lane(2018{\natexlab{a}})}]{liu2018adversarial}
Bing Liu and Ian Lane. 2018{\natexlab{a}}.
\newblock Adversarial learning of task-oriented neural dialog models.
\newblock \emph{SIGDIAL}, pages 350--359.

\bibitem[{Liu and Lane(2018{\natexlab{b}})}]{liu2018end}
Bing Liu and Ian Lane. 2018{\natexlab{b}}.
\newblock End-to-end learning of task-oriented dialogs.
\newblock In \emph{NAACL: Student Research Workshop}, pages 67--73.

\bibitem[{Mesnil et~al.(2015)Mesnil, Dauphin, Yao, Bengio, Deng, Hakkani-Tur,
  He, Heck, Tur, Yu et~al.}]{mesnil2015using}
Gr{\'e}goire Mesnil, Yann Dauphin, Kaisheng Yao, Yoshua Bengio, Li~Deng, Dilek
  Hakkani-Tur, Xiaodong He, Larry Heck, Gokhan Tur, Dong Yu, et~al. 2015.
\newblock Using recurrent neural networks for slot filling in spoken language
  understanding.
\newblock \emph{IEEE/ACM Transactions on Audio, Speech, and Language
  Processing}, 23(3):530--539.

\bibitem[{Oh and Rudnicky(2000)}]{oh2000stochastic}
Alice~H Oh and Alexander~I Rudnicky. 2000.
\newblock Stochastic language generation for spoken dialogue systems.
\newblock In \emph{Proceedings of the 2000 NAACL Workshop on Conversational
  systems-Volume 3}, pages 27--32. Association for Computational Linguistics.

\bibitem[{Papangelis et~al.(2018)Papangelis, Papadakos, Stylianou, and
  Tzitzikas}]{papangelis2018spoken}
Alexandros Papangelis, Panagiotis Papadakos, Yannis Stylianou, and Yannis
  Tzitzikas. 2018.
\newblock Spoken dialogue for information navigation.
\newblock In \emph{SIGDIAL}, pages 229--234. Association for Computational
  Linguistics.

\bibitem[{Papineni et~al.(2002)Papineni, Roukos, Ward, and
  Zhu}]{papineni2002bleu}
Kishore Papineni, Salim Roukos, Todd Ward, and Wei-Jing Zhu. 2002.
\newblock Bleu: a method for automatic evaluation of machine translation.
\newblock In \emph{Proceedings of the 40th annual meeting on association for
  computational linguistics}, pages 311--318. Association for Computational
  Linguistics.

\bibitem[{Reiter and Dale(2000)}]{reiter2000building}
Ehud Reiter and Robert Dale. 2000.
\newblock \emph{Building natural language generation systems}.
\newblock Cambridge university press.

\bibitem[{Rieser and Lemon(2010)}]{rieser2010natural}
Verena Rieser and Oliver Lemon. 2010.
\newblock Natural language generation as planning under uncertainty for spoken
  dialogue systems.
\newblock In \emph{EMNLP}, pages 105--120. Springer.

\bibitem[{Rojas-Barahona et~al.(2016)Rojas-Barahona, Gasic, Mrksic, hao Su,
  Ultes, Wen, and Young}]{RojasBarahona2016ExploitingSA}
Lina~Maria Rojas-Barahona, Milica Gasic, Nikola Mrksic, Pei hao Su, Stefan
  Ultes, Tsung-Hsien Wen, and Steve~J. Young. 2016.
\newblock Exploiting sentence and context representations in deep neural models
  for spoken language understanding.
\newblock In \emph{COLING}.

\bibitem[{Schatzmann et~al.(2007)Schatzmann, Thomson, Weilhammer, Ye, and
  Young}]{schatzmann2007agenda}
Jost Schatzmann, Blaise Thomson, Karl Weilhammer, Hui Ye, and Steve Young.
  2007.
\newblock Agenda-based user simulation for bootstrapping a pomdp dialogue
  system.
\newblock In \emph{Human Language Technologies 2007: The Conference of the
  North American Chapter of the Association for Computational Linguistics;
  Companion Volume, Short Papers}, pages 149--152. Association for
  Computational Linguistics.

\bibitem[{Serban et~al.(2016)Serban, Sordoni, Bengio, Courville, and
  Pineau}]{serban2016building}
Iulian~Vlad Serban, Alessandro Sordoni, Yoshua Bengio, Aaron~C Courville, and
  Joelle Pineau. 2016.
\newblock Building end-to-end dialogue systems using generative hierarchical
  neural network models.
\newblock In \emph{AAAI}, volume~16, pages 3776--3784.

\bibitem[{Shah et~al.(2018)Shah, Hakkani-Tur, Liu, and
  Tur}]{shah2018bootstrapping}
Pararth Shah, Dilek Hakkani-Tur, Bing Liu, and Gokhan Tur. 2018.
\newblock Bootstrapping a neural conversational agent with dialogue self-play,
  crowdsourcing and on-line reinforcement learning.
\newblock In \emph{NAACL, Volume 3 (Industry Papers)}, volume~3, pages 41--51.

\bibitem[{Shang et~al.(2015)Shang, Lu, and Li}]{P15-1152}
Lifeng Shang, Zhengdong Lu, and Hang Li. 2015.
\newblock \href {https://doi.org/10.3115/v1/P15-1152} {Neural responding
  machine for short-text conversation}.
\newblock In \emph{ACL-IJCNLP}, pages 1577--1586.

\bibitem[{Singh et~al.(1999)Singh, Kearns, Litman, and
  Walker}]{singh2000reinforcement}
Satinder~P. Singh, Michael~J. Kearns, Diane~J. Litman, and Marilyn~A. Walker.
  1999.
\newblock Reinforcement learning for spoken dialogue systems.
\newblock In \emph{NIPS}, pages 956--962. The {MIT} Press.

\bibitem[{Stent et~al.(2005)Stent, Marge, and Singhai}]{stent2005evaluating}
Amanda Stent, Matthew Marge, and Mohit Singhai. 2005.
\newblock Evaluating evaluation methods for generation in the presence of
  variation.
\newblock In \emph{CICLing}, volume 3406 of \emph{Lecture Notes in Computer
  Science}, pages 341--351. Springer.

\bibitem[{Su et~al.(2017)Su, Budzianowski, Ultes, Gasic, and
  Young}]{su2017sample}
Pei{-}Hao Su, Pawel Budzianowski, Stefan Ultes, Milica Gasic, and Steve~J.
  Young. 2017.
\newblock Sample-efficient actor-critic reinforcement learning with supervised
  data for dialogue management.
\newblock \emph{SIGDIAL}, pages 147--157.

\bibitem[{Sutskever et~al.(2014)Sutskever, Vinyals, and Le}]{Sutskever:2014}
Ilya Sutskever, Oriol Vinyals, and Quoc~V. Le. 2014.
\newblock Sequence to sequence learning with neural networks.
\newblock In \emph{NIPS}, pages 3104--3112.

\bibitem[{Tur and Mori(2011)}]{SLUBook}
G.~Tur and R.~De Mori, editors. 2011.
\newblock \emph{Spoken Language Understanding: Systems for Extracting Semantic
  Information from Speech}.
\newblock John Wiley and Sons, New York, NY.

\bibitem[{T{\"{u}}r et~al.(2012)T{\"{u}}r, Deng, Hakkani{-}T{\"{u}}r, and
  He}]{Tr2012TowardsDU}
G{\"{o}}khan T{\"{u}}r, Li~Deng, Dilek Hakkani{-}T{\"{u}}r, and Xiaodong He.
  2012.
\newblock Towards deeper understanding: Deep convex networks for semantic
  utterance classification.
\newblock In \emph{ICASSP}, pages 5045--5048. {IEEE}.

\bibitem[{Walker et~al.(2002)Walker, Rambow, and Rogati}]{walker2002training}
Marilyn~A Walker, Owen~C Rambow, and Monica Rogati. 2002.
\newblock Training a sentence planner for spoken dialogue using boosting.
\newblock \emph{Computer Speech \& Language}, 16(3-4):409--433.

\bibitem[{Wen et~al.(2015)Wen, Gasic, Mrksic, Su, Vandyke, and
  Young}]{wen2015semantically}
Tsung{-}Hsien Wen, Milica Gasic, Nikola Mrksic, Pei{-}hao Su, David Vandyke,
  and Steve~J. Young. 2015.
\newblock Semantically conditioned lstm-based natural language generation for
  spoken dialogue systems.
\newblock \emph{EMNLP}, pages 1711--1721.

\bibitem[{Wen et~al.(2017)Wen, Vandyke, Mrksic, Gasic, Rojas-Barahona, Su,
  Ultes, and Young}]{wen2016network}
Tsung-Hsien Wen, David Vandyke, Nikola Mrksic, Milica Gasic, Lina~M
  Rojas-Barahona, Pei-Hao Su, Stefan Ultes, and Steve Young. 2017.
\newblock A network-based end-to-end trainable task-oriented dialogue system.
\newblock \emph{EACL}, pages 438--449.

\bibitem[{Williams et~al.(2017)Williams, Asadi, and Zweig}]{williams2017hybrid}
Jason~D Williams, Kavosh Asadi, and Geoffrey Zweig. 2017.
\newblock Hybrid code networks: practical and efficient end-to-end dialog
  control with supervised and reinforcement learning.
\newblock In \emph{ACL (Volume 1: Long Papers)}, volume~1, pages 665--677.

\bibitem[{Wu et~al.(2018)Wu, Li, Liu, Gao, and Yang}]{wu2018switch}
Yuexin Wu, Xiujun Li, Jingjing Liu, Jianfeng Gao, and Yiming Yang. 2018.
\newblock Switch-based active deep dyna-q: Efficient adaptive planning for
  task-completion dialogue policy learning.
\newblock \emph{arXiv preprint arXiv:1811.07550}.

\bibitem[{Xu and Sarikaya(2013)}]{Xu2013ConvolutionalNN}
Puyang Xu and Ruhi Sarikaya. 2013.
\newblock Convolutional neural network based triangular {CRF} for joint intent
  detection and slot filling.
\newblock \emph{ASRU}, pages 78--83.

\bibitem[{Yin et~al.(2016)Yin, Jiang, Lu, Shang, Li, and
  Li}]{Yin:2016:NGQ:3060832.3061037}
Jun Yin, Xin Jiang, Zhengdong Lu, Lifeng Shang, Hang Li, and Xiaoming Li. 2016.
\newblock Neural generative question answering.
\newblock In \emph{IJCAI'16}, pages 2972--2978. AAAI Press.

\bibitem[{Young et~al.(2013)Young, Ga{\v{s}}i{\'c}, Thomson, and
  Williams}]{young2013pomdp}
Steve Young, Milica Ga{\v{s}}i{\'c}, Blaise Thomson, and Jason~D Williams.
  2013.
\newblock Pomdp-based statistical spoken dialog systems: A review.
\newblock \emph{Proceedings of the IEEE}, 101(5):1160--1179.

\bibitem[{Zen et~al.(2009)Zen, Tokuda, and Black}]{zen2009statistical}
Heiga Zen, Keiichi Tokuda, and Alan~W. Black. 2009.
\newblock Statistical parametric speech synthesis.
\newblock \emph{Speech Communication}, 51(11):1039--1064.

\bibitem[{Zhang and Wang(2016)}]{Zhang2016AJM}
Xiaodong Zhang and Houfeng Wang. 2016.
\newblock A joint model of intent determination and slot filling for spoken
  language understanding.
\newblock In \emph{IJCAI}.

\end{thebibliography}
\bibliographystyle{acl_natbib}
\end{footnotesize}

\appendix
\section{Example Dialogues}

Here we show two example dialogues generated by the \emph{WoLF-PHC} agents.

\begin{table*}[h]
    \centering
    \begin{tabular}{|r|l|}
        \hline
        Seeker goal & Constr(food=italian), Constr(pricerange=cheap), Req(addr), Req(phone) \\ \hline \hline
Provider & hello , welcome to the cambridge restaurant System? \\ 
& you can ask for restaurants by area , price range or food type . \\
Seeker & cheap restaurant  \\
Provider & what part of town do you have in mind?  \\
Seeker & italian food  \\
Provider & pizza hut city centre is a nice restaurant in the centre of town serving italian food  \\
Seeker & phone number  \\
Provider & the phone number is 01223 323737 .  \\
Seeker & address  \\
Provider & sure , it is on regent street city centre  \\
Seeker & (no output) \\
Provider & what part of town do you have in mind? \\
Seeker & good bye  \\
Provider & good bye  \\
\\ \hline
    \end{tabular}
    \caption{A good example dialogue.}
    \label{tab:example}
    \vspace{-2mm}
\end{table*}

\begin{table*}[h]
    \centering
    \begin{tabular}{|r|l|}
        \hline
        Seeker goal & Constr(food=african), Constr(pricerange=expensive), Req(addr) \\ \hline \hline
Provider & hello , welcome to the cambridge restaurant System? how may i help you? \\
Seeker & expensive restaurant  \\
Provider & what part of town do you have in mind?  \\
Seeker & african food  \\
Provider & bedouin is an expensive restaurant in the centre of town serving african food  \\
Seeker & i do not care  \\
Provider & sorry i am a bit confused ; please tell me again what you are looking for .  \\
Seeker & repeat  \\
Provider & what part of town do you have in mind?  \\
Seeker & address  \\
Provider & sure , it is on 100 mill road city centre  \\
Seeker & good bye  \\
Provider & good bye  \\ \hline
    \end{tabular}
    \caption{An imperfect example dialogue.}
    \label{tab:example}
    \vspace{-2mm}
\end{table*}

\end{document}